\documentclass[sigconf]{acmart}
\AtBeginDocument{%
  \providecommand\BibTeX{{%
    \normalfont B\kern-0.5em{\scshape i\kern-0.25em b}\kern-0.8em\TeX}}}

\copyrightyear{2023} 
\acmYear{2023} 
\setcopyright{acmlicensed}\acmConference[UIST '23]{The 36th Annual ACM Symposium on User Interface Software and Technology}{October 29-November 1, 2023}{San Francisco, CA, USA}
\acmBooktitle{The 36th Annual ACM Symposium on User Interface Software and Technology (UIST '23), October 29-November 1, 2023, San Francisco, CA, USA}
\acmPrice{15.00}
\acmDOI{10.1145/3586183.3606790}
\acmISBN{979-8-4007-0132-0/23/10}

\usepackage{acmart-taps}

\usepackage{xspace,xpunctuate}
\newcommand{\ie}{{i.e.,}\xspace}
\newcommand{\cf}{{cf.}\xspace}
\newcommand{\eg}{{e.g.,}\xspace}
\newcommand{\ea}{{et~al\xperiod}\xspace}

\newcommand{\feedback}[2][]{\emph{``#2''} (#1)}
\newcommand{\practitioner}{visualization engineer\xspace}
\newcommand{\sigmacomp}{Sigma Computing} 
\newcommand{\system}{VegaProf\xspace}
\newcommand{\vegaplus}{VegaPlus\xspace} 
\newcommand{\systemLink}{\href{https://github.com/cagataydemiralp/vegaprof}{GitHub (https://github.com/cagataydemiralp/vegaprof)}}

\begin{document}

\title[\system]%
      {\system: Profiling Vega Visualizations\vspace{-15pt}}

\author{Junran Yang}
\authornote{Work done while at Sigma Computing.}
\email{junran@cs.washington.edu}
\orcid{0000-0002-8467-2917}
\affiliation{%
  \institution{University of Washington}
  \city{Seattle}
  \state{Washington}
  \country{USA}
}

\author{Alex B{\"a}uerle}
\authornotemark[1] 
\email{alex@a13x.io}
\orcid{0000-0003-3886-8799}
\affiliation{%
  \city{San Francisco}
  \state{California}
  \country{USA}}

\author{Dominik Moritz}
\email{domoritz@cmu.edu}
\orcid{0000-0002-3110-1053}
\affiliation{%
  \institution{Carnegie Mellon University}
  \city{Pittsburgh}
  \state{Pennsylvania}
  \country{USA}
}

\author{Çağatay Demiralp}
\authornotemark[1] 
\email{cagatay@csail.mit.edu}
\orcid{0009-0003-2080-0443}
\affiliation{%
  \institution{MIT CSAIL}
  \city{Cambridge}
  \state{Massachusetts}
  \country{USA}}



\begin{abstract}
Domain-specific languages (DSLs) for visualization aim to facilitate visualization creation by providing abstractions that offload implementation and execution details from users to the system layer.
Therefore, DSLs often execute user-defined specifications by transforming them into intermediate representations (IRs) in successive lowering operations. 

However, DSL-specified visualizations can be difficult to profile and, hence, optimize due to the layered abstractions.
To better understand visualization profiling workflows and challenges, we conduct formative interviews with visualization engineers who use Vega in production.
Vega is a popular visualization DSL that transforms specifications into dataflow graphs, which are then executed to render visualization primitives.
Our formative interviews reveal that current developer tools are ill-suited for visualization profiling since they are disconnected from the semantics of Vega's specification and its IRs at runtime.

To address this gap, we introduce \system{}, the first performance profiler for Vega visualizations.
\system{} instruments the Vega library by associating a declarative specification with its compilation and execution.
Integrated into a Vega code playground, \system{} coordinates visual performance inspection at three abstraction levels: function, dataflow graph, and visualization specification.
We evaluate \system{} through use cases and feedback from visualization engineers as well as original developers of the Vega library.
Our results suggest that \system makes visualization profiling more tractable and actionable by enabling users to interactively probe time performance across layered abstractions of Vega.
Furthermore, we distill recommendations from our findings and advocate for co-designing visualization DSLs together with their introspection tools.
\end{abstract}

\ccsdesc[500]{Human-centered computing~User interface toolkits}

\keywords{Visualization; DSL; Profiler; Debugger; Dataflow; Developer tools}

\begin{teaserfigure}
 \includegraphics[width=\linewidth]{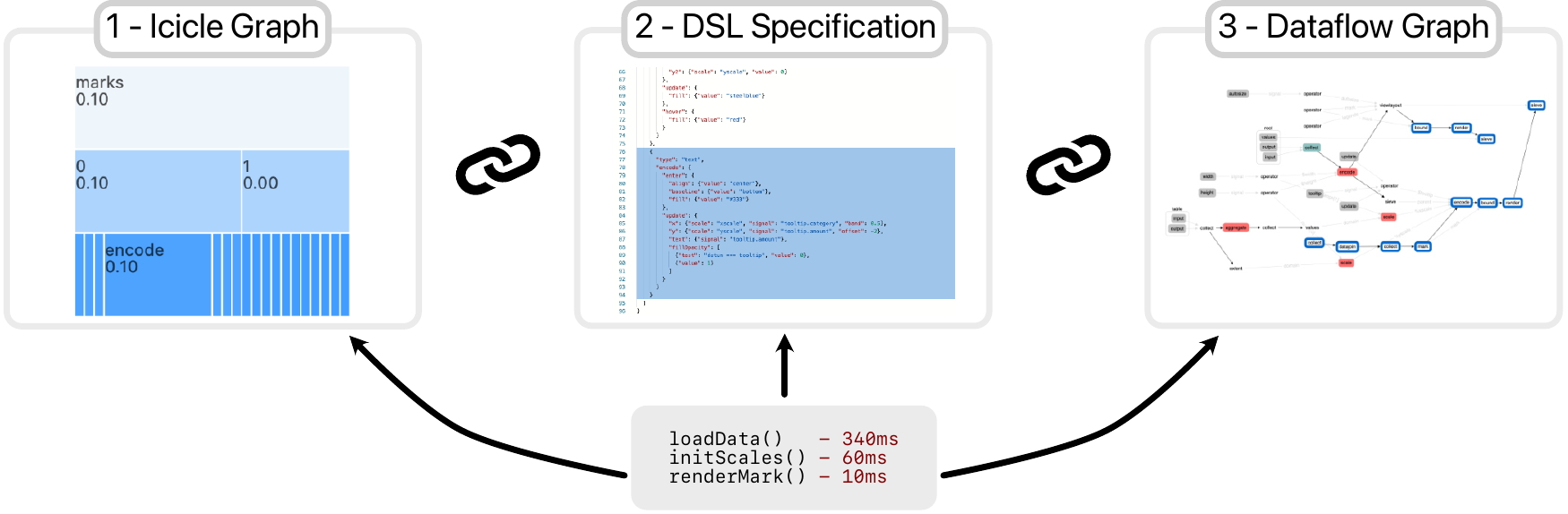}  
  \caption{{
  \system{} records low-level execution times and encodes them in coordinated visualizations corresponding to different abstraction levels of Vega's DSL execution. 
  (1) An icicle graph provides an overview of where most time is spent.
  (2) Unlike previous debugging approaches, which relied on trial-and-error-based changes of the visualization specification, \system{} maps performance measures directly to the Vega specification.
  (3) To help analyze performance on different levels, \system{} augments the dataflow graph with performance measures.
  }}
  \label{fig:teaser}
\end{teaserfigure}


\maketitle

\section{Introduction}\label{sec:intro} 

Domain-specific languages (DSLs) such as Vega~\cite{vega} and gglpot2~\cite{ggplot} enable both end-users and system developers~\cite{mcnutt2023no} to rapidly explore the visualization design space using abstractions that simplify coding and defer low-level control flow to the system layer~\cite{Satyanarayan2016reactive}. 
For example, Vega can parse a visualization specification written in JSON, where the underlying schema is informed by the grammar of graphics~\cite{wilkinson2012grammar}, into a dataflow graph abstracting the required computations and their dependencies; the resulting dataflow graph is then used to execute low-level visualization operations, including rendering functions.
This successive creation of intermediate representations (IRs) to compile and execute DSL code is called the \emph{lowering process}~\cite{beischl2021profiling}. 


However, what makes DSLs useful (\eg{} abstracting away implementation and execution details) can also make them challenging to understand~\cite{Naimipour2020engaging} and debug~\cite{hoffswell2015debugging, hoffswell2018augmenting}.
Users of visualization DSLs have limited visibility to code execution~\cite{hoffswell2015debugging}.
Nevertheless, they still need to ensure that their visualizations are \textit{responsive} and \textit{scalable}.
Visualizations include many elements affecting performance, including layout, encoding, data transformation, and rendering.
Due to the intricate dependency between these elements and ever-increasing data sizes, users may spend considerable iterative effort on performance optimizations, which makes profiling and introspection of visualization specifications challenging without adequate tools.
On the other hand, browser profilers cannot provide a multi-level performance trace~\cite{chromedev} as they don’t have access to mappings between IRs of a visualization DSL.
As general profilers for web applications, they are inadequate for DSLs, which require affordances to reason about lower-level abstractions in terms of higher-level ones. For example, users might know which rendering function slows down a visualization, but this might not help them to locate relevant parts in the DSL or its resulting dataflow graph.
As a result, to address performance problems, practitioners often rely on trial-and-error procedures, which can be error-prone, time-consuming, and taxing. 

In response, we introduce \system{} (\autoref{fig:teaser}) to enable easy and effective time-performance profiling for Vega visualizations.
\system{} instruments the Vega library to record low-level execution times and encodes and annotates them in coordinated visualizations corresponding to different abstraction levels of Vega execution.
\system{} enables users to trace performance from the code segments of a visualization specification to the dataflow graph and the executed functions and back through interactive visual inspection within a familiar developer tool, the Vega Editor.

\system{} is the first performance profiler for a visualization DSL, which contributes to tooling for visualization development and has synergistic connections to broader systems research~\cite{beischl2021profiling} and dataflow DSLs.
Although implemented in the context of Vega, the presented techniques generalize to other DSLs that instantiate dataflow systems.
For example, DBMSs optimize SQL queries and generate dataflow graphs of operators for execution. 

While the architecture of VegaProf, connecting layers of abstractions, is similar to prior interactive systems at a high level, its design contributions lie in anchoring the design explicitly on a DSL's hierarchical IRs and bidirectionally coupling them via visual interaction.
This combination offers an effective (semantically direct~\cite{hutchins1985direct}) and conceptually straightforward recipe for designing future interactive introspection tools for DSLs with rich design variations possible. 
Scalability remains a crucial problem for data visualization in practice, particularly in cloud data analytics.
Thus, future DSLs will need to incorporate constructs for optimizations.
Therefore, performance profilers for visualization DSLs will be even more critical in the future, for which VegaProf sets the stage. 

We inform the design of \system{} with a formative study involving three visualization engineers.
We subsequently evaluate its value and usability for users through three use cases and a summative study with five visualization engineers and two of the original Vega developers.
Both studies are conducted with participants who have real stakes in performance, providing added support for the validity of the results.
Our findings demonstrate the utility and usability of \system{} and our profiling framework, reinforce the effectiveness of encoding performance measurements in an overview visualization (an icicle graph) linked to the underlying visualization specification, and establish that different abstractions appeal to different user groups.
We release \system{} on \systemLink{} as open-source software to support future research and applications.

\section{Related Work}\label{sec:related} 

\system builds upon prior research on visualization debugging and dataflow system profiling.

\subsection{Debugging Visualizations}

Data visualization research has a long history of investigating DSLs for visualization specification (\eg \cite{wilkinson2012grammar,hanrahan2006vizql,wickham2010layered,bostock2011d3,satyanarayan2016vega,wills2017brunel}), but research into linting and debugging visualizations is nascent.
McNutt and Kindlmann~\cite{mcnutt2018linting} introduce a visualization linter that checks a predefined set of rules on a given visualization and returns a list of failed rules with explanations; this postprocessing approach is disconnected from the development workflow and does not localize errors for rendered visualizations in their specifications.
In contrast, VisuaLint~\cite{hopkins2020visualint} annotates visualizations in situ with red marks; these marks, akin to conventional linting-error visualizations in IDEs, cannot be traced back to the visualization specification.
To rectify defective visualization designs, VizLinter~\cite{chen2021vizlinter} highlights flaws directly in the visualization specification: it maps flaws to DSL code while suggesting potential fixes. 
Since interactions can be particularly challenging to debug, Hoffswell \ea~\cite{hoffswell2016visual} propose debugging techniques designed for reactive visualizations; to provide needed detail to the \practitioner{}, they track state through interactions that are mapped to a visual debugging interface.
However, they use signal names to map problems to the Vega specification, so their technique works only to debug interactions and is thus not generally applicable.

This prior work focuses on errors in the visualization specification rather than performance problems.
It thus targets a different problem space than \system, which enables interactive time performance profiling. 

\subsection{Profiling Dataflow Systems}
Vega is a form of dataflow system~\cite{veen1986dataflow}.
Bidirectional coupling of the visualization specification (code) with its associated dataflow graph and rendering functions is central to interactive profiling in \system.
Dataflow graphs are a common abstraction used by myriad tools across domains beyond data visualization (e.g., PyTorch, TensorFlow, Spark, Flink, Naiad, SQL).
Earlier work presents profiling tools to help discover performance issues in dataflow systems~\cite{gathani2020debugging}.
For example, Perfopticon~\cite{moritz2015perfopticon} shows the runtime distribution of individual query operators and per-worker execution traces; like our approach, it also maps the profiling result to user input and leverages the importance of connecting high-level abstractions with execution in profiling.
Battle~\ea disentangle SQL queries as a series of intermediate queries to help developers debug the behavior of their queries in StreamTrace~\cite{battle2016making}.
Similarly, Grust \ea~\cite{grust2011true} link intermediate query results to the SQL code that generated them instead of using representative visualizations.
Mapping performance directly to code has been a common paradigm in recent research~\cite{cito2019interactive}.
However, none of the earlier approaches target visualization DSLs or consider dataflow graphs as a profiling entity.
Beischl \ea~\cite{beischl2021profiling} propose a multi-level performance profiling technique specifically for dataflow-based systems.
We adopt a similar approach for developing an interactive profiler for Vega, building our primary visualizations on different IRs of Vega and their interactive coordination through brushing and linking.
\system complements this work~\cite{beischl2021profiling} by leveraging the bidirectional maps between DSL IRs to improve visual interactive profiling.

\section{Formative Interviews}\label{sec:formative} 

Our work was initially motivated by our efforts to help developers improve visualization performance in an interactive data analytics product~\cite{gale2022sigma}. 


To assess the needs of visualization developers, we interviewed three professional \practitioner{}s for whom the programmatic generation of Vega visualizations is part of their daily work.
Though our interviewees all worked with Vega specifications on a daily basis, they had varying levels of Vega expertise.
P1 and P2 regarded Vega as the configuration mechanism for visualizations but had less knowledge of the underlying execution; in contrast, P3 had a basic understanding of Vega internals, such as how a specification instantiates its underlying dataflow graph.

We conducted and recorded our semi-structured interviews via online video conferencing software. 
We covered a list of predefined topics and questions, leaving time for open discussion.
Topics discussed included their performance optimization needs, current practice and tools, and desired performance profiling features, and we present our findings for each topic below.

\noindent\textbf{Performance optimization.}
We discovered that performance issues \feedback[P1]{usually have large impacts} because interviewees' customers get frustrated with slow system responsiveness.
However, interviewees acknowledged that they do not proactively monitor or conduct regular testing for performance but deal reactively with performance issues after customers report them.
They attributed this to the fact that performance issues are currently difficult to localize, debug, and fix.
Indeed, interviewees underlined that \feedback[P2]{performance issues are often neglected} and \feedback[P1]{once we find the causes, we often tell customers not to perform such operations}.
Furthermore, \practitioner{}s typically \feedback[P3]{limit the data input to a Vega spec to less than 25k [data points] to prevent a lot of slow rendering issues}.
Moving forward, these stopgap solutions are not sustainable since interviewees \feedback[P1]{have seen a lot of questions and requests regarding visualization performance}.

\noindent\textbf{Current practice and tools.}
To reason about poor performance, \practitioner{}s typically \feedback[P1]{have Zoom meetings with customers to talk about problematic visualizations}.
Then, they often \feedback[P1 and P3]{simulate the configurations} and test them in a sandbox environment.
As such, they lack specialized tooling for performance debugging, instead relying on \feedback[All participants]{the Vega Editor in combination with Chrome's devtools}.
However, the problem with this is that \feedback[P1]{devtools can only tell you that the issues are caused by Vega function calls, but it can’t help with locating them inside the specification}.
Interviewees therefore often relied on their previous experience and had no tooling support to test their hypotheses. 

\noindent\textbf{Desired features.}
When asked about what features could help overcome their problems, interviewees asked for a \feedback[P3]{breakdown of the transforms, mark rendering, etc., that can immediately indicate which lines [in the DSL] caused the issue}.

\subsection{Design Goals}\label{sec:criteria} 

Based on our interview observations and our broader conversations with visualization practitioners, we distilled the following requirements for a Vega performance profiler:

\noindent\textbf{Bidirectional timing mapping.}
To help \practitioner{}s discover the root cause of performance bottlenecks, it is not sufficient simply to measure function execution times.
Instead, to make informed decisions for performance optimization, practitioners must understand how different parts of their code specification incur performance costs and how they are grounded in Vega's IRs, including the associated dataflow graph and function calls.

\noindent\textbf{Multi-level profiling insights.}
Given a bidirectional mapping of timings of the DSL's IRs, practitioners must be able to investigate mapping results.
Interactive visualizations that surface timing measurements can facilitate such introspection, enabling practitioners to trace and contextualize performance measurements in situ through the IRs used in Vega's lowering pipeline. 
Users can target different abstraction levels depending on their expertise and the nature of their work. 

\noindent\textbf{Familiar development environment.}
We aim to support \practitioner{}s in a familiar environment, aligning the profiler with their existing workflows and making it approachable.
As with most software users, \practitioner{}s adopt new tools only if the burden of entry does not outweigh their benefits.
Therefore, we chose to integrate the profiler into the Vega Editor~\cite{vegaeditor}, which is widely used by practitioners, reducing 
the friction for adoption.        

\begin{figure}
    \centering
    \includegraphics[width=\linewidth]{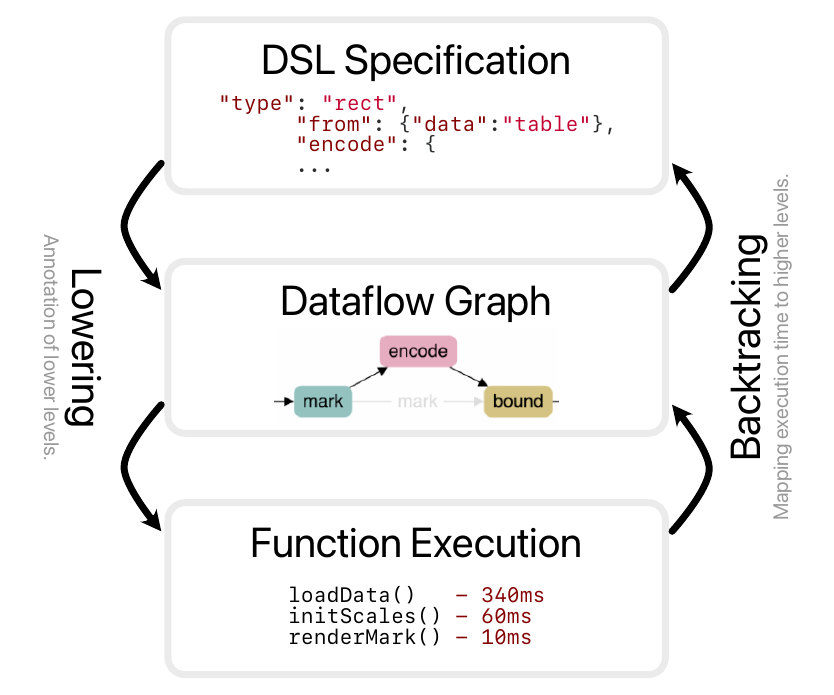}
    \caption{
During the lowering process, the user-defined DSL specification is parsed into a computation graph and then into functions to be evaluated. We add annotations during the lowering process and then trace profiling results back to higher levels of abstraction.}
    \label{fig:mapping}
\end{figure}
\begin{figure*}
    \centering
    \includegraphics[width=\linewidth]{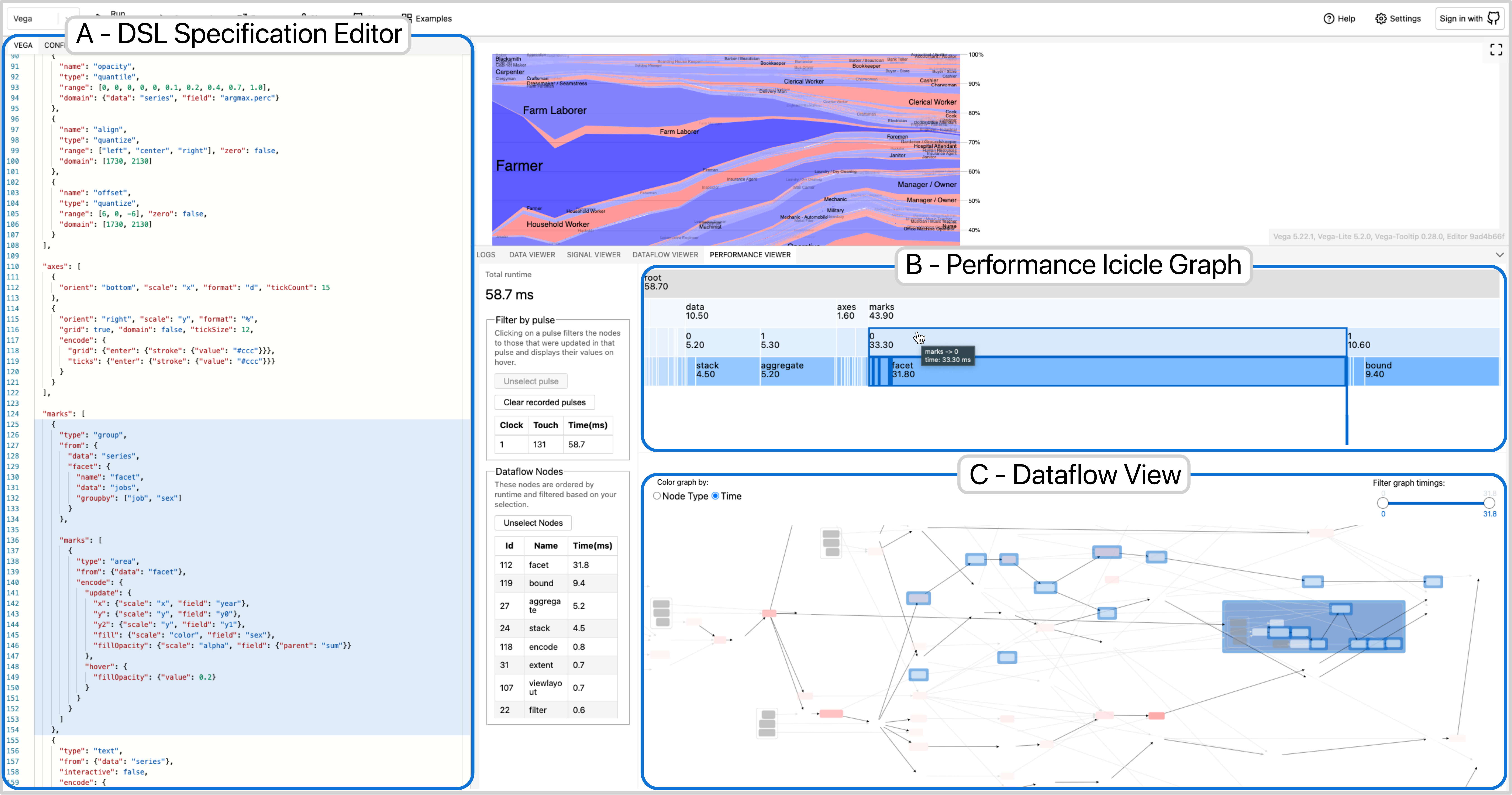}
    \caption{
    \system{}'s visual inspection functionalities are implemented in a familiar development environment---the Vega Editor.
    (A) We highlight the selected regions of DSL code when inspecting performance bottlenecks.
    (B) An icicle graph depicts the measurement of rendering function execution time.
    (C) We highlight the nodes contributing to selected timing measurements in the Vega-generated dataflow graph.
    Note how hovering over the icicle graph highlights the corresponding elements in the dataflow graph and DSL editor.
    }
    \label{fig:application}
\end{figure*}

\section{Visualization profiling}
To bridge the gap between visualization specification and low-level execution, we implement a bidirectional profiling map that tracks the association between higher and lower abstraction levels as they undergo code generation and compilation.
Then, we visually represent the collected information and integrate it as coordinated views in the Vega Editor.

\subsection{Bidirectional Profiling Map}\label{sec:profiling_map}

The recording of function execution times is well-established for time-profiling.
However, effective profiling instruments for a visualization DSL rely on a bidirectional mapping of execution time measurements and DSL segments with semantic meanings.
Only with a mapping as shown in \autoref{fig:mapping} can \practitioner{}s analyze performance bottlenecks in the context of the IRs that result from  DSL transformations.

To realize such a mapping, we annotate the nodes of the \emph{dataflow graph description} as they are created when parsing the visualization specification.
Specifically, since a Vega specification is a nested JSON object, we store (1) the path of keys to access a specification component as the \textit{key} in our map and (2) the list of nodes instantiated by it during the lowering process as the \textit{value}. 
This way, we can reverse this mapping, associating dataflow graph nodes with corresponding lines of DSL code.

Once the dataflow graph description is transformed into a \emph{dataflow runtime}, where the nodes represent functions to be evaluated, we further annotate these functions with the respective dataflow graph nodes to realize such a mapping for this second lowering process.
Our measurements of function execution time can therefore be mapped back to the node of the dataflow graph that triggered the execution.
By chaining the annotations from the two phases, we can assign the execution of individual functions not only to nodes in the dataflow graph but also to lines and blocks of DSL code.
Using this bidirectional mapping, execution times can be traced from the function level back to the highest level of operation, namely the DSL specification for the visualization.

This approach directly addresses our first design goal of creating a bidirectional map between Vega's IRs.
While we use this inter-IR indexing approach only to improve profiling instruments, it could also be helpful for other introspection tools, such as learning about Vega's lowering process or dataflow debugging.
Furthermore, our approach could be expanded to other libraries in the Vega ecosystem. For example, Vega-Lite specifications compile to Vega, which could be viewed as one more level of abstraction in our method. 
In theory, this bidirectional profiling map can be visualized and analyzed in any environment.

Our addition to the Vega source code is minimal and lightweight.
The main implementation effort went into integrating the profiling results as coordinated views into the Vega Editor, which we describe next.

\subsection{Visual Performance Inspection}

Based on the information obtained from the bidirectional mapping of profiling results, we provide a visual interface that lets \practitioner{}s take action to improve performance of their visualization designs.
We implemented this interactive performance profiling interface as an extension to the Vega Editor to provide an environment familiar to  \practitioner{}s.
The Vega Editor, a well-established visualization development tool used by many Vega users, is implemented as a live playground that runs in the browser using the Monaco Editor that powers Visual Studio Code for specification input.
Our profiling map is updated whenever users modify their specification.
In this way, the Vega Editor provides bidirectional performance tracing with links across DSL line numbers, specification blocks, dataflow nodes, and function execution times
We use React~\cite{React} and Redux~\cite{redux} for \system's interface in the Vega Editor; for its visualizations, we use D3~\cite{bostock2011d3} and Cytoscape.js~\cite{cytoscape}. 

A new performance tab provides a performance icicle graph (\autoref{fig:application} (B)) and augments the dataflow graph (\autoref{fig:application} (C)) and the DSL specification editor (\autoref{fig:application} (A)).
These three components are connected through brushing and linking techniques using our bidirectional profiling map (\autoref{sec:profiling_map}) as the underlying data source.
We thus map selections, \emph{mouseover} events, and zoom transitions that occur in one of the three views to the other two.
Such interactions are indicated using blue highlights; hovered over items are assigned a semi-transparent blue highlight, and selected items are highlighted in full blue consistently across all visualizations.

In addition to these main views, the Vega Editor further displays the visualization that results from the provided specification.
If the visualization is interactive, the resulting profiling and operator states from interaction events are recorded as multiple \emph{pulses}.
The first pulse marks the initial rendering of the visualization, and subsequent ones are added whenever the visualization is updated based on user interaction.
Pulses are selected from the pulse table, and the icicle and dataflow graphs are updated to show only the operators being re-evaluated and their timings.
In addition, pulses augment the dataflow graph via node tooltips by providing insights into data changes in the individual nodes along every pulse.
By default, we show profiling results for the initial rendering pulse for both static and interactive charts; in so doing, \practitioner{}s can use our visualizations and profiling results not only to debug the initial rendering process but also to improve interaction performance.
Directly above the pulse selection, we prominently show the total runtime of the selected pulse so users have an anchor to put all visual interface timings into context.
    
\subsubsection{DSL Specification Editor}

The DSL specification editor is prominently positioned at the left edge of the Vega Editor (\cf{}~\autoref{fig:application} (A)), marking a natural entry point for debugging.
It represents the highest level of abstraction for \system, directly connecting performance profiles to the DSL code that defines the visualization.
Since this level can be directly influenced by \practitioner{}s, it is often where their time-performance analysis begins.

To map function execution times to blocks of the DSL specification, we consider different levels of ranges in the specification.
These blocks directly map to JSON's hierarchical object structure in the Vega specification.
For example, a user would specify both the \emph{x-axis} and \emph{y-axis} blocks under the \emph{axis} block. 
In Vega, these blocks are the units that \practitioner{}s would associate with visual components.
Hence, this is the level at which they would make edits, \eg{} changing the type of mark used for rendering or how data is mapped to these marks.
As such, this way of clustering parts of the DSL naturally aligns with how Vega users understand and modify the specifications.

Hovering over one of these blocks of DSL code highlights the respective specification segment.
This gives \practitioner{}s insight into how much time individual blocks of the visualization specification require during rendering.
Clicking on such a highlighted code block selects it  for further inspection.
As noted earlier, highlights and selections are transferred to the corresponding elements in the dataflow and performance icicle graphs.
More importantly, we also implemented the reverse linking directions from the icicle and dataflow graphs. 
For example, investigating elements that require significant rendering time in the icicle graph scrolls users to the corresponding code segment of the DSL and highlights it.
As a result, one can easily interpret the high-level responsibility of a particular element of the icicle chart and the dataflow graph by inspecting the highlighted block in the specification. 

This connection makes such performance measures insightful and actionable since \practitioner{}s can directly adjust relevant parts of the DSL code.

\begin{figure*}
    \centering
    \includegraphics[width=\linewidth]{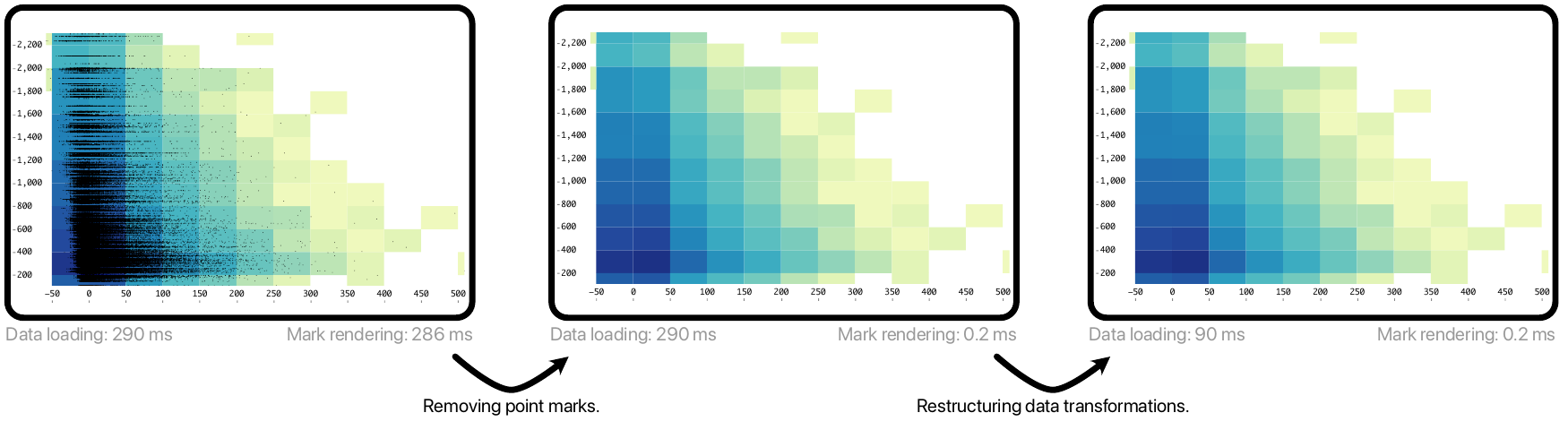}
    \caption{
    We use a scatter plot with binned aggregation as an example during our evaluation. The rendering time is about 600 ms initially. The individual point marks do not add substantial value to the visualization while obstructing the heatmap.
    Thus, removing them does not undermine the message the visualization aims to communicate and reduces the mark rendering time almost to zero.
    Furthermore, the data transformation is specified suboptimally, requiring Vega to copy the data.
    Restructuring the data transformation further saves  about 200 ms without changing the visualization.
    }
    \label{fig:study_spec}
\end{figure*}

\subsubsection{Dataflow View}

As the first IR of the Vega visualization grammar, the DSL specification the \practitioner{} provides is transformed into a dataflow graph.
In the dataflow view, our visual performance inspection interface enables analysis at a more detailed level.
This view (\autoref{fig:application} (C)) contains a visualization of the parsed dataflow graph that results from Vega's DSL transformation. 
It can be color-coded by node type or, more conveniently for performance analysis, by node runtime.
We use D3's \emph{interpolateReds} color scale to encode node runtime since red is often used as an alarm color, drawing attention to the most expensive nodes during rendering.

Whenever a node is selected from this graph visualization, the dataflow graph gets transformed to show only the subgraph with connections to the selected node based on dependency.
A zoom-in animation further highlights selected nodes.
If a selection comes from any other visualization, such as the DSL editor or the performance icicle graph, we analyze the nodes involved in the selected subset of performance analysis elements and employ filtering and zooming similar to that used for direct node selection.
This interaction concept further embraces the combined analysis of Vega's different IRs, similar to how interaction with the DSL specification editor is mapped to all other visualizations.

To directly identify the most time-consuming nodes, \system further includes a table of all nodes positioned next to the dataflow graph.
This table, based on the dataflow graph, is ordered by node execution time, placing the most performance-intensive nodes at the top.
This tabular visualization can thus be used as an entry-point of the analysis on the dataflow level since it guides the \practitioner{}'s attention directly to nodes of interest.

\subsubsection{Performance Icicle Graph}

Positioned directly above the dataflow graph, the performance icicle graph (\cf{}~\autoref{fig:application} (B)) functions as an intermediate representation between the dataflow graph and the DSL specification editor.
It is defined by its different levels of aggregation, from coarse performance elements to more fine-grained ones.
To symbolize this aggregation structure, we color coarser levels in gray and light blue and fine-grained levels in dark blue.
The most detailed level in this icicle graph directly represents nodes of the dataflow graph; however, the icicle graph also visualizes the hierarchical structure of the Vega DSL at its higher levels, connecting the two other views in one visualization.

Hovering over and selecting elements in the icicle graph works the same  as it does in our other visualizations.
The icicle graph additionally zooms into selected elements to provide more detailed information about a selection.
Like the dataflow graph, this zooming and highlighting might also be triggered by events from other visualizations.
In sum, the icicle graph, with its different levels of performance aggregation and linked interaction concepts, serves as a bridge between the specification editor and the dataflow view.

\section{Use-Cases}

This section describes how \system can help \practitioner{}s discover and resolve performance problems of their Vega visualizations through three example use-cases.
These cases highlight how connecting different IRs can help to debug performance and, specifically, how directly linking performance bottlenecks to Vega's specification makes such analyses actionable.

\subsection{Visualization Design Decisions}\label{subsec:heatmap}

Mary is a \practitioner{} in the data analysis team of a large airline.
She wants to analyze the effect of flight distance on the delay of flights based on a dataset that contains information on three million flights.
She considers a scatter plot to visualize the data. When she specifies the scatter plot in Vega, she notices that the visualization she has created is too slow to be usable.

With \system, Mary loads her data and visualization specification into the Vega Editor and analyzes the performance of her visualization.
Through an investigation of the icicle Graph, she immediately notices that mark rendering consumes most of the total visualization generation time. Looking at the connected location in the visualization specification, she notices that rendering individual scatter marks for millions of flights is too slow to sustain interactivity.
Since Mary is seeking general trends rather than individual flight information, she decides to remove these marks and instead render a heatmap for binned results.
\system helps users make design decisions by illustrating tradeoffs between performance and visualization design.

Next, Mary notices that loading the data was relatively slow.
Using the dataflow graph, she locates an operation that copies part of the data during the transformation stage. Since the relevant part of the Vega specification is highlighted when she hovers over the corresponding dataflow node, she identifies the problem and modifies the transformation code to make data processing more performant.

After these modifications, Mary further notices that while much faster than before, data processing remains her chief performance bottleneck.
The final step she could take is to pre-aggregate data instead of binning it on demand.
However, since the data frequently changes, she decides against this approach and accepts the initial loading time because of the data transformation that makes her specification more portable. 

\begin{figure*}
    \centering
    \includegraphics[width=\linewidth]{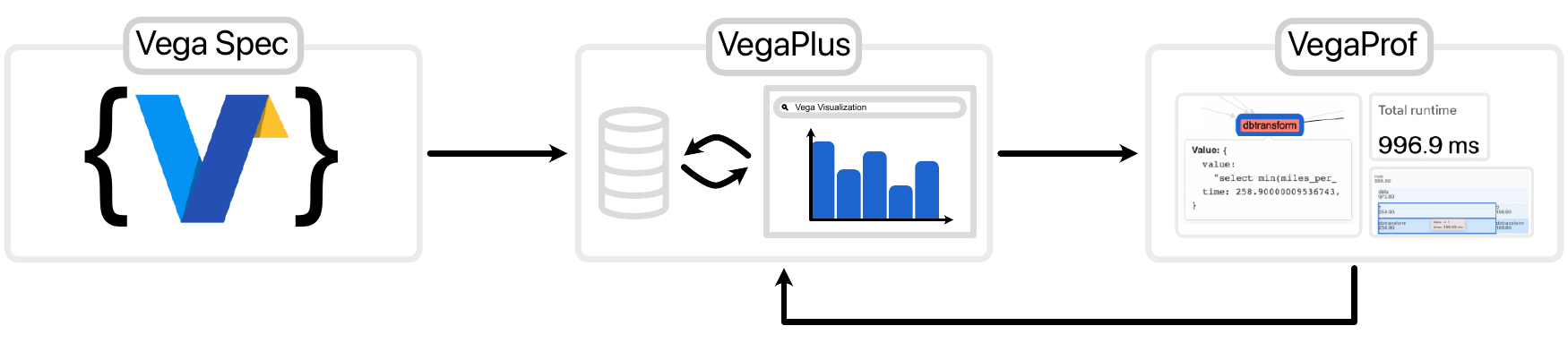}
    \caption{
    Amy transforms a raw Vega specification using \vegaplus to distribute computation across the backend and frontend.
    She inspects the resulting performance profile in \system to identify bottlenecks in her implementation.
    She uses the dataflow graph along with its associated tooltip, the icicle graph, and the total rendering time to gain performance insights about the execution plan generated by \vegaplus.
    Then, she interactively changes her implementation of \vegaplus to improve performance.
    }
    \label{fig:vegaplus}
\end{figure*}

\subsection{Offloading Expensive Operations} 

Alice works as a \practitioner{} for a software company.
Her team implements product features for UI-based visualization authoring. Their product lets users without programming expertise create various visualizations to explore data in a cloud data warehouse (CDW). 
They use Vega as the underlying technology to specify visualizations through their product's UI. 

By default, Vega requires all data to be loaded and processed in the client's browser.
However, it is computationally impossible to query the CDW for the raw data and transfer it to the browser's memory for processing in Vega.
Therefore, Alice decides to pre-process the data with SQL queries upon  request from the visualization so that the query result is ready to be directly mapped to visual channels without further Vega transforms.
However, her testing visualization is not sufficiently fast in its initial rendering and does not seamlessly respond to user interactions.

When Alice inspects \system's pulses and dataflow nodes affected by the pulses, she sees that each interaction triggers a request to the CDW, blocking the entire dataflow graph. 
After analyzing the requests, Alice notices that intermediate results can be cached to reduce the interaction latency.
She rewrites the query to fetch those results only once and moves the downstream transformations into the Vega specification.
As a result, time-consuming computations are offloaded to the backend while operations with interactive latency are executed in the browser. 

\subsection{High-Performance Tool Development}

Amy is developing \vegaplus~\cite{yang2022demonstration}, a system to automatically offload computationally expensive Vega operations for large-scale data.
Given a Vega specification, \vegaplus partitions data operations between  the server and client.
Amy uses \system to help her with different development tasks, as shown in ~\autoref{fig:vegaplus}.

To optimize Vega's execution, she needs to understand how Vega instantiates the dataflow graph.
Without \system, she had to log the dataflow object, go through the nested structure in a browser console, and meticulously inspect each level.
Therefore, locating the dataflow graph section she needs to inspect required substantial work.
With \system, Amy uses the visual exploration interface to inspect the holistic dataflow graph structure and variables inside each node.
This helps her rapidly locate nodes of interest and complements the previous manual inspection method. 

\vegaplus automatically rewrites operations as SQL queries.
At runtime, interaction signals pass the parameters to query builders to generate executable queries.
Amy selects a pulse from \system's pulse table and hovers over the transform node to check the queries sent to the DBMS and their results.
Therefore, she need not manually monitor and log interaction information. 

Finally, \system helps Amy profile and visualize the performance of \vegaplus in detail before running benchmarks.
An execution plan that performs well for initial rendering may suffer from significant interaction latency.
Since users are often more tolerant of slower initial rendering if it benefits faster interactions, Amy wants \vegaplus to optimize for interaction performance.
To prototype various execution plans, Amy uses \system to simulate interactions and then analyzes the different execution plans with the help of \system's performance chart and pulse table.

\section{Interview Study}

To further evaluate \system, we conducted a qualitative user study with five \practitioner{}s at \sigmacomp{} and two Vega developers who work on its toolkit. In the following, we provide details on the setup of our study, including our participant pool and the procedure and data used.

\noindent\textbf{Participants.}
We selected five visualization engineers who develop visualizations for hundreds of analysts across different companies; their visualizations must be scalable and perform well.They work with Vega on a daily basis, although they had different levels of background knowledge of the system's internal dataflow; as such, they are  representative of  \system's target audience. 
The two Vega developers we interviewed develop and maintain the Vega toolkit. They are experts with the Vega internals. 

\noindent\textbf{Procedure.}
To identify the benefits and limitations of \system, we held a 30-minute think-aloud session with each participant individually.
We first gave a quick tutorial of \system during these sessions before participants experimented with the profiler.
For these experiments, our participants were given 20 minutes to access  \system with a visualization specification preloaded. 
They were asked to explore the profiler based on two guiding questions:  \emph{How could you adjust the specification for faster rendering?}
 \emph{How can the data processing be improved?}
We then encouraged them to share a specification from their recent work and show us how they would use \system to inspect it. 
Finally, we asked our study participants to rate different aspects of \system on a five-point Likert scale via an online questionnaire.

\noindent\textbf{Data and specification.}
The specification we used for this evaluation renders a scatter plot overlaid on a binned heatmap (\cf{} \autoref{fig:study_spec}), as described in \autoref{subsec:heatmap}.
Input data to the visualization consists of three million rows, each rendering one data point.
While a user interacts with the chart by panning and zooming actions, the dataflow graph re-calculates the bins and aggregated values.
We selected this specification since rendering or constantly aggregating a large number of data points is prone to performance issues.

\section{Results} 
We now present the main findings of our evaluation and discuss their implications.

\subsection{Visualization Engineers}
We first report on the qualitative feedback elicited from our sessions with five \practitioner{}s.

With the help of the icicle graph, all participants identified point marks as the most time-consuming components.
They located the relevant part in the specification by hovering on the icicle graph, which they found helpful: \feedback[P3]{It is impressive that you can highlight the spec [from the icicle chart].}. 
Based on the highlighted region in the specification, all participants recognized that a scatter plot might not be ideal, both visually and computationally, and removed it.
The visual connection between performance measures and  relevant parts of the visualization specification through highlighting was one of the most well-received features of \system.
It was seen as a substantial improvement over the current way of debugging slow specifications, which is to \feedback[P1]{just guess which part is the cause and modify it to see if it solves the issue or not}.
One participant stressed the importance of the icicle graph to their workflow, since without \system \feedback[P5]{we could separate the data transform out and profile it programmatically, but there was no way to do that for everything else, [including] the marks, the rendering, etc.}.

As participants inspected the resulting performance after this first edit, they discovered that the runtime for rendering the marks was significantly reduced.
Subsequently, they recognized that, with the modified specification, the most time-consuming operations were the transformations used to process the data.
Participants were able to select the relevant dataflow nodes responsible for the performance bottleneck.
However, most of them lacked the background knowledge about how nodes are instantiated from the specification through parsing and compilation.
Specifically, they were not able to infer from the node name \emph{relay} that an unnecessary data copy operation caused the performance bottleneck.
P5 managed to solve the task by removing unnecessary operations, although doing so required a hint from us: \feedback[P5]{Now that you told me a transform in the spec can be expanded to multiple operations, I can see it in the icicle chart, and everything makes sense to me}.
P1 and P2 could not find a solution that addressed this performance bottleneck. 
After we explained how to reconstruct the data transformation pipeline, they acknowledged that knowing Vega internals and inspecting the dataflow graph would help optimize  specification authoring: \feedback[P1]{I'm surprised that doing this can save so much [execution] time!}.

In general, we observed that participants spent most of their time exploring the icicle graph \feedback[P2]{because it exactly tells you what part of rendering is taking up all the time}.
From there, they typically inspected the highlighted segments of the specification.
Participants spent less time inspecting the dataflow graph.
Even those who previously debugged many Vega visualizations with the dataflow graph initially found the icicle chart more helpful than the dataflow graph. 
They intensified their focus on the dataflow graph only after being reminded of the connection between it and the icicle graph.  
In part, this might be because most participants lacked understanding of node names and \feedback[P2]{so far have just been using Vega as a black box [...], assuming that it would work well}.
However, they also mentioned that \feedback[P4]{the connection between the spec and dataflow graph, and the structural features [in the dataflow graph], could be really helpful to understand what goes on behind the scenes for Vega}.

After the guided exploration, three participants asked to directly explore specifications they recently worked on in \system. 
We observed how they used \system to validate or reject their assumptions about a given specification.
P2 showed us a scatter plot with categorical data, admitting initial surprise 
that \feedback[P2]{the axes took the longest to render, and then the marks were comparatively shorter [...] that's not what I would have guessed initially}; P2 then realized that their dataset was relatively small, while the categorical variables mapped to the axes had a high cardinality.
P4 shared with us a Sankey diagram they had been working on for \sigmacomp{}'s product. During development, they frequently inspected the dataflow graph to understand and debug the connection between nodes.
Concluding that \feedback[P4]{I'm not surprised that the ``linkpath'' and ``datajoin'' operations took the most time}, they verified that performance conformed with their mental model for such a diagram. 
Finally, P5 wanted to explore a visualization automatically generated from a Sigma Workbook~\cite{gale2022sigma}.
They exported the specification of a simple chart to test their mental model of how it was implemented and found that the Dashboard framework implemented it well: \feedback[P5]{It's so fast that the axes take half of the rendering time.}.

\begin{figure}
    \centering
    \includegraphics[width=\linewidth]{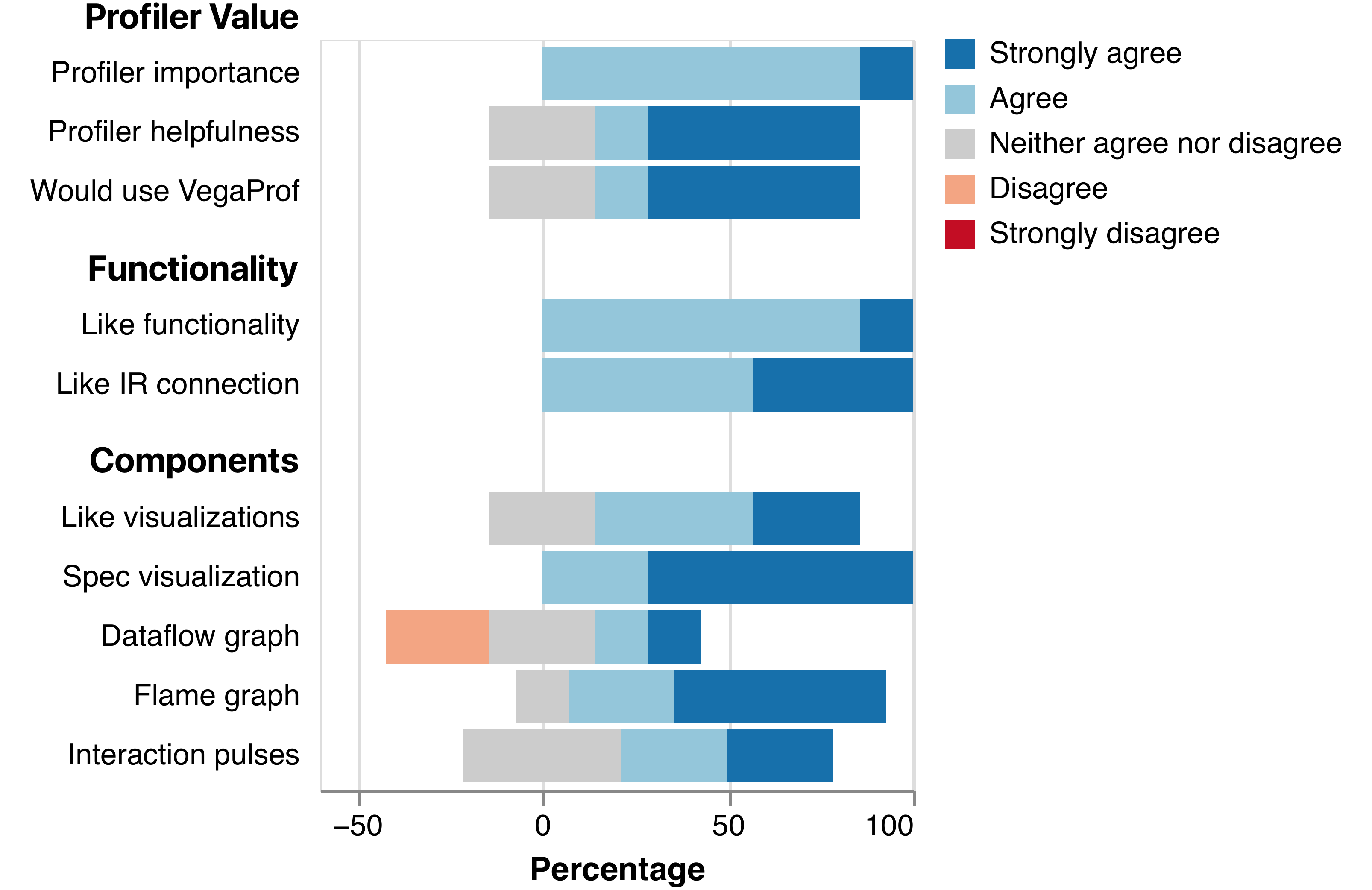}
    \caption{Most participants wanted to use \system{} for profiling Vega visualizations in their daily work. 
    They also agreed on the value of all but one \system{} visualization components. Preferred primarily by the two Vega developers, the dataflow graph visualization received mixed feedback, which would benefit from further investigation. All participants expressed significant support for the utility of \system{} functionalities, including tracing performance through a DSL’s IR and visualizing these traces for performance profiling.
    }
    \label{fig:ratings}
\end{figure}

\subsection{Vega System Developers}
The two Vega developers we interviewed are experts in Vega's internals. 
P6 is a maintainer of the Vega repository and designs and develops software that is built on top of Vega. 
P7 is one of Vega's main designers and developers.

First, both interviewees went through the specification to understand the example we provided.
Each immediately pointed out the point marks as the culprit. 
P6 also proposed further options to improve the visualization design;  
in addition to removing all point marks, they tested different data sampling parameters to achieve a balance between acceptable latency and a representative result. 
Like the group of \practitioner{}s, these more advanced Vega users  found the icicle chart and its linkage to the specification most efficient for performance debugging: \feedback[P6]{The icicle graph shows me what the expensive things are and then seeing that link, it's nice to be able to see the part in the spec.}.

Both  noticed that the data transformation operations were taking as much time as the mark rendering. 
Looking at the icicle graph, both identified that in the \emph{relay} transformations, an internal data copying transform was the source of the problem.
P7 reasoned more than P6 about whether they should keep the relay: it is slow but necessary if there are downstream filtering operations because it copies and caches the data only once, and the effort is amortized by future interactions. 
However, by selecting interaction pulses, P7 confirmed that his hypothesis was not the case, and the relay was indeed redundant; hence, they decided to remove it by rewriting the specification. 
They used the dataflow graph \feedback[P7]{to confirm some things that I already knew, but without the icicle graph I probably wouldn't have, certainly not as quickly, realized that the relay was causing us a significant hit}.
P7 acknowledged that their position as a Vega developer gave them an advantage and that \feedback[P7]{I don't know if other people would know that}, remarking that the dataflow graph could serve as more of an expert tool: \feedback[P7]{I do think this is an expert level thing where someone has to be very knowledgable about Vega internals.}.

P6 wondered \feedback[P6]{whether [...] Vega could be better at} automatically optimizing the pipeline rather than relying on users to handle these kinds of optimizations. 
On the other hand, P7 maintained that any modification depends on \feedback[P7]{who the chart is for, and what are they trying to learn and decide}.

After reviewing the specification we presented, P7 was interested in exploring a visualization with three histograms connected by a brushing and linking interaction. 
They focused on profiling the interactions by inspecting \system's pulse table and suggested providing a better overview and segmentation of the pulses to increase readability: \eg \feedback[P7]{I would like to segment pulses on initiating event type, running time, and/or recency}.


\subsection{Post-Study Questions}

All participants, both Vega developers and \practitioner{}s, provided  feedback on their experience and takeaways from our interview session on a 5-point Likert scale. The chart displays sentiments towards ten questions as percentages, with neutral responses straddling the 0\% mark.
The results of this evaluation are shown in~\autoref{fig:ratings}.

In general, participants found visualization profilers useful.
Furthermore, most of them would like to use \system for their visualization profiling needs, affirming the utility of our profiler implementation.
This confirms the findings from our formative interviews.
Our participants also appreciated how we coordinated Vega's intermediate representations via visual interaction, supporting our architectural design choices.
While they especially appreciated \system's linking of different visual elements, they were divided on the usefulness of interactive pulses.
The most controversial of the visualizations was the dataflow graph; 
while some found it valuable, others did not see it as beneficial for their workflows.

\noindent\subsection{Study Limitations}
Our study design was limited by the small number of participants and their similar background within user groups. 
In particular, the visualization engineers at \sigmacomp{} work primarily on static charts and were less familiar with the concept of ``pulses.'' 
Thus, they focused more on the initial rendering during the study.
On the other hand, Vega system developers showed more interest in profiling user interactions. 
An evaluation with a broader audience for a longer period of time would likely uncover more insights, especially with respect to the usage of the pulse table for interactive visualizations.

\section{Discussion and Recommendations}

\system was rated positively by our study participants, confirming the importance of this work, its architectural design choices, and most of the visualization decisions.
We were intrigued by the motivation of our participants to try \system on their own Vega specifications.
Their ability to make immediate sense of its performance profiles and view performance measurements in the light of their DSL code indicates that \system is applicable beyond our evaluation scope.

\noindent\textbf{Performance tracing.}
All participants heavily used and commented positively about the connection of performance measures between Vega's IRs.
Highlighting parts of the specification especially helped Vega users and developers locate and fix performance bottlenecks, establishing that tracing performance measures through a DSL's IRs helps improve time-performance profiling.

\noindent\textbf{DSL design.}
Considering our experience with the internal data copying operation (\ie relay) and the feedback of Vega developers, we believe that visualization DSLs could be further optimized.
To improve both the system developer and end-user experience, we advocate for jointly designing future visualization DSLs together with their introspection tools, such as debuggers and profilers.
Furthermore, recommendations for performance improvements could be included in such tools, similar to related work in the machine learning domain~\cite{schoop2021umlaut,yu2020skyline}, to help users with  profiling.

\noindent\textbf{Profiling tool audience.}
While the dataflow graph is one of the main debugging components in the Vega Editor to date, \practitioner{}s found the icicle graph, which we added to \system, much more helpful.
This both underscores the importance of our icicle graph visualization but also raises questions about the utility of the dataflow graph.
One participant noted that \feedback[P3]{We use a lot of the Chrome devtools, and their entire interface is basically only the icicle chart}, attributing their focused view partly to previous habits.
In addition, one of the Vega developers remarked that the dataflow graph could be more of an expert tool.

Further research on making dataflow systems more understandable, including explanations of individual nodes and meaningful clustering, might help \practitioner{}s make better use of the dataflow graph.
Additionally, we deem it essential to consider the target audience of an introspection tool since different types of users need different visualizations.

\section{Future Work}\label{sec:discussion} 

While \system is the first profiler for the Vega visualization grammar, we hope that future work can further refine the user experience with visualization  profiling, as outlined below.

\noindent\textbf{Direct visualization connection.}
Naimipour~\ea ~\cite{Naimipour2020engaging} call for better support to aid program understanding for visualization DSLs. 
With \system, the rendered visualization does not display profiling information.
However, by expanding performance tracking to the scene graph, components of the resulting visualization could be linked to \system's profiling visualizations. 
By adding this additional layer to the profiler, a connection between the specification and the rendered visualization could be established to help further trace performance issues.
This would be valuable both for DSL understanding and learning.

\noindent\textbf{Profiling dataflow systems.}
While \system focuses on the Vega DSL, its underlying approach and visual interaction design could readily apply to other DSLs.
Many DSLs use a similar lowering process and dataflow graph as an intermediate representation.
Further generalization of our approach could enable performance profiling for a wide array of these systems, further broadening the availability of accessible profiling.

\noindent\textbf{Performance at scale.}
Our evaluation shows how our approach improves performance for individual visualizations.
However, DSLs are also frequently used for visualization generation at scale.
As a result, thousands of users could benefit from automatically generated visualizations, \eg in web applications under different browser and cloud configurations.
Future research could extend this work to help \practitioner{}s profile the distributed performance of visualizations. 

\section{Conclusion}\label{sec:conclusion}
We introduced \system, an interactive profiler enabling in-depth analysis of performance bottlenecks in Vega visualizations. We based its design on formative interviews surfacing the difficulty of Vega visualization performance debugging. \system replaces inefficient performance debugging practices characterized by trial and error procedures. It brings visual profiling affordances to Vega's IRs by hooking into the lowering process, enabling the display of profiling results directly on the dataflow graph and visualization specification.  We demonstrated \system in action through three use cases and evaluated using feedback elicited from five \practitioner{}s and two Vega developers. In our evaluation, participants successfully located and brainstormed about the ways to address performance bottlenecks. 

While our work here focused on a DSL for visualization design, DSLs are ubiquitous across domains and can invariably benefit from better introspection support to help with myriad tasks ranging from debugging and profiling to system teaching. Future research can extend the applications of the techniques operationalized by \system{}; in particular, the instrumentation for bidirectional mapping and the corresponding visual interaction design that couples visualizations of IRs annotated with measures of interest can benefit developer tools for DSLs at large. 
%


\bibliographystyle{ACM-Reference-Format}
\bibliography{main}


\begin{thebibliography}{34}


\ifx \showCODEN    \undefined \def \showCODEN     #1{\unskip}     \fi
\ifx \showDOI      \undefined \def \showDOI       #1{#1}\fi
\ifx \showISBNx    \undefined \def \showISBNx     #1{\unskip}     \fi
\ifx \showISBNxiii \undefined \def \showISBNxiii  #1{\unskip}     \fi
\ifx \showISSN     \undefined \def \showISSN      #1{\unskip}     \fi
\ifx \showLCCN     \undefined \def \showLCCN      #1{\unskip}     \fi
\ifx \shownote     \undefined \def \shownote      #1{#1}          \fi
\ifx \showarticletitle \undefined \def \showarticletitle #1{#1}   \fi
\ifx \showURL      \undefined \def \showURL       {\relax}        \fi
\providecommand\bibfield[2]{#2}
\providecommand\bibinfo[2]{#2}
\providecommand\natexlab[1]{#1}
\providecommand\showeprint[2][]{arXiv:#2}

\bibitem[Abramov and Clark(2015)]%
        {redux}
\bibfield{author}{\bibinfo{person}{Dan Abramov} {and} \bibinfo{person}{Andrew
  Clark}.} \bibinfo{year}{2015}\natexlab{}.
\newblock \bibinfo{title}{Redux}.
\newblock \bibinfo{howpublished}{\url{https://redux.js.org/}}.
\newblock
\newblock
\shownote{Accessed on April 3, 2023}.


\bibitem[Battle et~al\mbox{.}(2016)]%
        {battle2016making}
\bibfield{author}{\bibinfo{person}{Leilani Battle}, \bibinfo{person}{Danyel
  Fisher}, \bibinfo{person}{Robert DeLine}, \bibinfo{person}{Mike Barnett},
  \bibinfo{person}{Badrish Chandramouli}, {and} \bibinfo{person}{Jonathan
  Goldstein}.} \bibinfo{year}{2016}\natexlab{}.
\newblock \showarticletitle{Making sense of temporal queries with interactive
  visualization}. In \bibinfo{booktitle}{\emph{Proceedings of the 2016 CHI
  Conference on Human Factors in Computing Systems}}.
  \bibinfo{pages}{5433--5443}.
\newblock


\bibitem[Beischl et~al\mbox{.}(2021)]%
        {beischl2021profiling}
\bibfield{author}{\bibinfo{person}{Alexander Beischl}, \bibinfo{person}{Timo
  Kersten}, \bibinfo{person}{Maximilian Bandle}, \bibinfo{person}{Jana Giceva},
  {and} \bibinfo{person}{Thomas Neumann}.} \bibinfo{year}{2021}\natexlab{}.
\newblock \showarticletitle{Profiling dataflow systems on multiple abstraction
  levels}. In \bibinfo{booktitle}{\emph{Proceedings of the Sixteenth European
  Conference on Computer Systems}}. \bibinfo{pages}{474--489}.
\newblock


\bibitem[Bostock et~al\mbox{.}(2011)]%
        {bostock2011d3}
\bibfield{author}{\bibinfo{person}{Michael Bostock}, \bibinfo{person}{Vadim
  Ogievetsky}, {and} \bibinfo{person}{Jeffrey Heer}.}
  \bibinfo{year}{2011}\natexlab{}.
\newblock \showarticletitle{D$^3$ data-driven documents}.
\newblock \bibinfo{journal}{\emph{IEEE transactions on visualization and
  computer graphics}} \bibinfo{volume}{17}, \bibinfo{number}{12}
  (\bibinfo{year}{2011}), \bibinfo{pages}{2301--2309}.
\newblock


\bibitem[Chen et~al\mbox{.}(2021)]%
        {chen2021vizlinter}
\bibfield{author}{\bibinfo{person}{Qing Chen}, \bibinfo{person}{Fuling Sun},
  \bibinfo{person}{Xinyue Xu}, \bibinfo{person}{Zui Chen},
  \bibinfo{person}{Jiazhe Wang}, {and} \bibinfo{person}{Nan Cao}.}
  \bibinfo{year}{2021}\natexlab{}.
\newblock \showarticletitle{Vizlinter: A linter and fixer framework for data
  visualization}.
\newblock \bibinfo{journal}{\emph{IEEE transactions on visualization and
  computer graphics}} \bibinfo{volume}{28}, \bibinfo{number}{1}
  (\bibinfo{year}{2021}), \bibinfo{pages}{206--216}.
\newblock


\bibitem[Cito et~al\mbox{.}(2019)]%
        {cito2019interactive}
\bibfield{author}{\bibinfo{person}{J{\"u}rgen Cito}, \bibinfo{person}{Philipp
  Leitner}, \bibinfo{person}{Martin Rinard}, {and} \bibinfo{person}{Harald~C
  Gall}.} \bibinfo{year}{2019}\natexlab{}.
\newblock \showarticletitle{Interactive production performance feedback in the
  IDE}. In \bibinfo{booktitle}{\emph{2019 IEEE/ACM 41st International
  Conference on Software Engineering (ICSE)}}. IEEE, \bibinfo{pages}{971--981}.
\newblock


\bibitem[Editor(2022)]%
        {vegaeditor}
\bibfield{author}{\bibinfo{person}{Vega Editor}.}
  \bibinfo{year}{2022}\natexlab{}.
\newblock \bibinfo{booktitle}{\emph{Vega Editor}}.
\newblock
\urldef\tempurl%
\url{https://vega.github.io/editor/}
\showURL{%
\tempurl}


\bibitem[Facebook(2013)]%
        {React}
\bibfield{author}{\bibinfo{person}{Facebook}.} \bibinfo{year}{2013}\natexlab{}.
\newblock \bibinfo{title}{React}.
\newblock \bibinfo{howpublished}{\url{https://reactjs.org/}}.
\newblock
\newblock
\shownote{Accessed on April 3, 2023}.


\bibitem[Gale et~al\mbox{.}(2022)]%
        {gale2022sigma}
\bibfield{author}{\bibinfo{person}{James Gale}, \bibinfo{person}{Max Seiden},
  \bibinfo{person}{Deepanshu Utkarsh}, \bibinfo{person}{Jason Frantz},
  \bibinfo{person}{Rob Woollen}, {and} \bibinfo{person}{{\c{C}}a{\u{g}}atay
  Demiralp}.} \bibinfo{year}{2022}\natexlab{}.
\newblock \showarticletitle{Sigma Workbook: A Spreadsheet for Cloud Data
  Warehouses}.
\newblock \bibinfo{journal}{\emph{arXiv preprint arXiv:2204.03128}}
  (\bibinfo{year}{2022}).
\newblock


\bibitem[Gathani et~al\mbox{.}(2020)]%
        {gathani2020debugging}
\bibfield{author}{\bibinfo{person}{Sneha Gathani}, \bibinfo{person}{Peter Lim},
  {and} \bibinfo{person}{Leilani Battle}.} \bibinfo{year}{2020}\natexlab{}.
\newblock \showarticletitle{Debugging database queries: A survey of tools,
  techniques, and users}. In \bibinfo{booktitle}{\emph{Proceedings of the 2020
  CHI Conference on Human Factors in Computing Systems}}.
  \bibinfo{pages}{1--16}.
\newblock


\bibitem[Grust et~al\mbox{.}(2011)]%
        {grust2011true}
\bibfield{author}{\bibinfo{person}{Torsten Grust}, \bibinfo{person}{Fabian
  Kliebhan}, \bibinfo{person}{Jan Rittinger}, {and} \bibinfo{person}{Tom
  Schreiber}.} \bibinfo{year}{2011}\natexlab{}.
\newblock \showarticletitle{True language-level SQL debugging}. In
  \bibinfo{booktitle}{\emph{Proceedings of the 14th International Conference on
  Extending Database Technology}}. \bibinfo{pages}{562--565}.
\newblock


\bibitem[Hanrahan(2006)]%
        {hanrahan2006vizql}
\bibfield{author}{\bibinfo{person}{Pat Hanrahan}.}
  \bibinfo{year}{2006}\natexlab{}.
\newblock \showarticletitle{VizQL: A Language for Query, Analysis and
  Visualization}. In \bibinfo{booktitle}{\emph{SIGMOD}}.
\newblock


\bibitem[Hoffswell et~al\mbox{.}(2015)]%
        {hoffswell2015debugging}
\bibfield{author}{\bibinfo{person}{Jane Hoffswell}, \bibinfo{person}{Arvind
  Satyanarayan}, {and} \bibinfo{person}{Jeffrey Heer}.}
  \bibinfo{year}{2015}\natexlab{}.
\newblock \showarticletitle{{Debugging Vega through Inspection of the Data Flow
  Graph}}. In \bibinfo{booktitle}{\emph{EuroVis Workshop on Reproducibility,
  Verification, and Validation in Visualization (EuroRV3)}},
  \bibfield{editor}{\bibinfo{person}{W.~Aigner},
  \bibinfo{person}{P.~Rosenthal}, {and} \bibinfo{person}{C.~Scheidegger}}
  (Eds.). \bibinfo{publisher}{The Eurographics Association}.
\newblock
\urldef\tempurl%
\url{https://doi.org/10.2312/eurorv3.20151144}
\showDOI{\tempurl}


\bibitem[Hoffswell et~al\mbox{.}(2016)]%
        {hoffswell2016visual}
\bibfield{author}{\bibinfo{person}{Jane Hoffswell}, \bibinfo{person}{Arvind
  Satyanarayan}, {and} \bibinfo{person}{Jeffrey Heer}.}
  \bibinfo{year}{2016}\natexlab{}.
\newblock \showarticletitle{Visual debugging techniques for reactive data
  visualization}. In \bibinfo{booktitle}{\emph{Computer Graphics Forum}},
  Vol.~\bibinfo{volume}{35}. Wiley Online Library, \bibinfo{pages}{271--280}.
\newblock


\bibitem[Hoffswell et~al\mbox{.}(2018)]%
        {hoffswell2018augmenting}
\bibfield{author}{\bibinfo{person}{Jane Hoffswell}, \bibinfo{person}{Arvind
  Satyanarayan}, {and} \bibinfo{person}{Jeffrey Heer}.}
  \bibinfo{year}{2018}\natexlab{}.
\newblock \showarticletitle{Augmenting Code with In Situ Visualizations to Aid
  Program Understanding}. In \bibinfo{booktitle}{\emph{Proceedings of the 2018
  CHI Conference on Human Factors in Computing Systems}} (Montreal QC, Canada)
  \emph{(\bibinfo{series}{CHI '18})}. \bibinfo{publisher}{Association for
  Computing Machinery}, \bibinfo{address}{New York, NY, USA},
  \bibinfo{pages}{1–12}.
\newblock
\showISBNx{9781450356206}
\urldef\tempurl%
\url{https://doi.org/10.1145/3173574.3174106}
\showDOI{\tempurl}


\bibitem[Hopkins et~al\mbox{.}(2020)]%
        {hopkins2020visualint}
\bibfield{author}{\bibinfo{person}{Aspen~K Hopkins}, \bibinfo{person}{Michael
  Correll}, {and} \bibinfo{person}{Arvind Satyanarayan}.}
  \bibinfo{year}{2020}\natexlab{}.
\newblock \showarticletitle{VisuaLint: Sketchy in situ annotations of chart
  construction errors}. In \bibinfo{booktitle}{\emph{Computer Graphics Forum}},
  Vol.~\bibinfo{volume}{39}. Wiley Online Library, \bibinfo{pages}{219--228}.
\newblock


\bibitem[Hutchins et~al\mbox{.}(1985)]%
        {hutchins1985direct}
\bibfield{author}{\bibinfo{person}{Edwin~L Hutchins}, \bibinfo{person}{James~D
  Hollan}, {and} \bibinfo{person}{Donald~A Norman}.}
  \bibinfo{year}{1985}\natexlab{}.
\newblock \showarticletitle{Direct manipulation interfaces}.
\newblock \bibinfo{journal}{\emph{Human--computer interaction}}
  \bibinfo{volume}{1}, \bibinfo{number}{4} (\bibinfo{year}{1985}),
  \bibinfo{pages}{311--338}.
\newblock


\bibitem[Inc.(2023)]%
        {chromedev}
\bibfield{author}{\bibinfo{person}{Alphabet Inc.}}
  \bibinfo{year}{2023}\natexlab{}.
\newblock \bibinfo{title}{Chrome Dev Tools}.
\newblock
  \bibinfo{howpublished}{\url{https://developer.chrome.com/docs/devtools/}}.
\newblock
\newblock
\shownote{Accessed on April 4, 2023}.


\bibitem[McNutt and Kindlmann(2018)]%
        {mcnutt2018linting}
\bibfield{author}{\bibinfo{person}{Andrew McNutt} {and} \bibinfo{person}{Gordon
  Kindlmann}.} \bibinfo{year}{2018}\natexlab{}.
\newblock \showarticletitle{Linting for visualization: Towards a practical
  automated visualization guidance system}. In
  \bibinfo{booktitle}{\emph{VisGuides: 2nd Workshop on the Creation, Curation,
  Critique and Conditioning of Principles and Guidelines in Visualization}}.
\newblock


\bibitem[McNutt(2023)]%
        {mcnutt2023no}
\bibfield{author}{\bibinfo{person}{Andrew~M. McNutt}.}
  \bibinfo{year}{2023}\natexlab{}.
\newblock \showarticletitle{No Grammar to Rule Them All: A Survey of JSON-style
  DSLs for Visualization}.
\newblock \bibinfo{journal}{\emph{IEEE Transactions on Visualization and
  Computer Graphics}} \bibinfo{volume}{29}, \bibinfo{number}{1}
  (\bibinfo{year}{2023}), \bibinfo{pages}{160--170}.
\newblock
\urldef\tempurl%
\url{https://doi.org/10.1109/TVCG.2022.3209460}
\showDOI{\tempurl}


\bibitem[Moritz et~al\mbox{.}(2015)]%
        {moritz2015perfopticon}
\bibfield{author}{\bibinfo{person}{Dominik Moritz}, \bibinfo{person}{Daniel
  Halperin}, \bibinfo{person}{Bill Howe}, {and} \bibinfo{person}{Jeffrey
  Heer}.} \bibinfo{year}{2015}\natexlab{}.
\newblock \showarticletitle{Perfopticon: Visual query analysis for distributed
  databases}. In \bibinfo{booktitle}{\emph{Computer Graphics Forum}},
  Vol.~\bibinfo{volume}{34}. Wiley Online Library, \bibinfo{pages}{71--80}.
\newblock


\bibitem[Naimipour et~al\mbox{.}(2020)]%
        {Naimipour2020engaging}
\bibfield{author}{\bibinfo{person}{Bahare Naimipour}, \bibinfo{person}{Mark
  Guzdial}, {and} \bibinfo{person}{Tamara Shreiner}.}
  \bibinfo{year}{2020}\natexlab{}.
\newblock \showarticletitle{Engaging Pre-Service Teachers in Front-End Design:
  Developing Technology for a Social Studies Classroom}. In
  \bibinfo{booktitle}{\emph{2020 IEEE Frontiers in Education Conference
  (FIE)}}. \bibinfo{pages}{1--9}.
\newblock
\urldef\tempurl%
\url{https://doi.org/10.1109/FIE44824.2020.9273908}
\showDOI{\tempurl}


\bibitem[Satyanarayan et~al\mbox{.}(2016a)]%
        {satyanarayan2016vega}
\bibfield{author}{\bibinfo{person}{Arvind Satyanarayan},
  \bibinfo{person}{Dominik Moritz}, \bibinfo{person}{Kanit Wongsuphasawat},
  {and} \bibinfo{person}{Jeffrey Heer}.} \bibinfo{year}{2016}\natexlab{a}.
\newblock \showarticletitle{Vega-lite: A grammar of interactive graphics}.
\newblock \bibinfo{journal}{\emph{IEEE transactions on visualization and
  computer graphics}} \bibinfo{volume}{23}, \bibinfo{number}{1}
  (\bibinfo{year}{2016}), \bibinfo{pages}{341--350}.
\newblock


\bibitem[Satyanarayan et~al\mbox{.}(2016b)]%
        {Satyanarayan2016reactive}
\bibfield{author}{\bibinfo{person}{Arvind Satyanarayan}, \bibinfo{person}{Ryan
  Russell}, \bibinfo{person}{Jane Hoffswell}, {and} \bibinfo{person}{Jeffrey
  Heer}.} \bibinfo{year}{2016}\natexlab{b}.
\newblock \showarticletitle{Reactive Vega: A Streaming Dataflow Architecture
  for Declarative Interactive Visualization}.
\newblock \bibinfo{journal}{\emph{IEEE Transactions on Visualization and
  Computer Graphics}} \bibinfo{volume}{22}, \bibinfo{number}{1}
  (\bibinfo{date}{jan} \bibinfo{year}{2016}), \bibinfo{pages}{659–668}.
\newblock
\showISSN{1077-2626}
\urldef\tempurl%
\url{https://doi.org/10.1109/TVCG.2015.2467091}
\showDOI{\tempurl}


\bibitem[Schoop et~al\mbox{.}(2021)]%
        {schoop2021umlaut}
\bibfield{author}{\bibinfo{person}{Eldon Schoop}, \bibinfo{person}{Forrest
  Huang}, {and} \bibinfo{person}{Bjoern Hartmann}.}
  \bibinfo{year}{2021}\natexlab{}.
\newblock \showarticletitle{Umlaut: Debugging deep learning programs using
  program structure and model behavior}. In
  \bibinfo{booktitle}{\emph{Proceedings of the 2021 CHI Conference on Human
  Factors in Computing Systems}}. \bibinfo{pages}{1--16}.
\newblock


\bibitem[Shannon et~al\mbox{.}(2003)]%
        {cytoscape}
\bibfield{author}{\bibinfo{person}{Paul Shannon}, \bibinfo{person}{Andrew
  Markiel}, \bibinfo{person}{Owen Ozier}, \bibinfo{person}{Nitin~S. Baliga},
  \bibinfo{person}{Jonathan~T. Wang}, \bibinfo{person}{Daniel Ramage},
  \bibinfo{person}{Nada Amin}, \bibinfo{person}{Benno Schwikowski}, {and}
  \bibinfo{person}{Trey Ideker}.} \bibinfo{year}{2003}\natexlab{}.
\newblock \showarticletitle{Cytoscape: A Community-Based Framework for Network
  Visualization and Analysis}.
\newblock \bibinfo{journal}{\emph{Genome Research}} \bibinfo{volume}{13},
  \bibinfo{number}{11} (\bibinfo{year}{2003}), \bibinfo{pages}{2498--2504}.
\newblock
\urldef\tempurl%
\url{https://doi.org/10.1101/gr.1239303}
\showDOI{\tempurl}


\bibitem[Veen(1986)]%
        {veen1986dataflow}
\bibfield{author}{\bibinfo{person}{Arthur~H Veen}.}
  \bibinfo{year}{1986}\natexlab{}.
\newblock \showarticletitle{Dataflow machine architecture}.
\newblock \bibinfo{journal}{\emph{ACM Computing Surveys (CSUR)}}
  (\bibinfo{year}{1986}).
\newblock


\bibitem[Vega(2022)]%
        {vega}
\bibfield{author}{\bibinfo{person}{Vega}.} \bibinfo{year}{2022}\natexlab{}.
\newblock \bibinfo{booktitle}{\emph{Vega \& Vega Lite Visualization Grammars}}.
\newblock
\urldef\tempurl%
\url{https://vega.github.io/}
\showURL{%
\tempurl}


\bibitem[Wickham(2010)]%
        {wickham2010layered}
\bibfield{author}{\bibinfo{person}{Hadley Wickham}.}
  \bibinfo{year}{2010}\natexlab{}.
\newblock \showarticletitle{A layered grammar of graphics}.
\newblock \bibinfo{journal}{\emph{Journal of Computational and Graphical
  Statistics}} (\bibinfo{year}{2010}).
\newblock


\bibitem[Wickham(2016)]%
        {ggplot}
\bibfield{author}{\bibinfo{person}{Hadley Wickham}.}
  \bibinfo{year}{2016}\natexlab{}.
\newblock \bibinfo{booktitle}{\emph{ggplot2: Elegant Graphics for Data
  Analysis}}.
\newblock \bibinfo{publisher}{Springer-Verlag New York}.
\newblock
\showISBNx{978-3-319-24277-4}
\urldef\tempurl%
\url{https://ggplot2.tidyverse.org}
\showURL{%
\tempurl}


\bibitem[Wilkinson(2012)]%
        {wilkinson2012grammar}
\bibfield{author}{\bibinfo{person}{Leland Wilkinson}.}
  \bibinfo{year}{2012}\natexlab{}.
\newblock \showarticletitle{The grammar of graphics}.
\newblock In \bibinfo{booktitle}{\emph{Handbook of computational statistics}}.
  \bibinfo{publisher}{Springer}, \bibinfo{pages}{375--414}.
\newblock


\bibitem[Wills(2018)]%
        {wills2017brunel}
\bibfield{author}{\bibinfo{person}{Graham Wills}.}
  \bibinfo{year}{2018}\natexlab{}.
\newblock \bibinfo{title}{Brunel v2.6}.
\newblock
  \bibinfo{howpublished}{https://github.com/Brunel-Visualization/Brunel}.
\newblock
\newblock
\shownote{Accessed: 2023-05-04}.


\bibitem[Yang et~al\mbox{.}(2022)]%
        {yang2022demonstration}
\bibfield{author}{\bibinfo{person}{Junran Yang}, \bibinfo{person}{Hyekang~Kevin
  Joo}, \bibinfo{person}{Sai~S Yerramreddy}, \bibinfo{person}{Siyao Li},
  \bibinfo{person}{Dominik Moritz}, {and} \bibinfo{person}{Leilani Battle}.}
  \bibinfo{year}{2022}\natexlab{}.
\newblock \showarticletitle{Demonstration of VegaPlus: Optimizing Declarative
  Visualization Languages}. In \bibinfo{booktitle}{\emph{Proceedings of the
  2022 International Conference on Management of Data}}.
  \bibinfo{pages}{2425--2428}.
\newblock


\bibitem[Yu et~al\mbox{.}(2020)]%
        {yu2020skyline}
\bibfield{author}{\bibinfo{person}{Geoffrey~X Yu}, \bibinfo{person}{Tovi
  Grossman}, {and} \bibinfo{person}{Gennady Pekhimenko}.}
  \bibinfo{year}{2020}\natexlab{}.
\newblock \showarticletitle{Skyline: Interactive In-Editor Computational
  Performance Profiling for Deep Neural Network Training}. In
  \bibinfo{booktitle}{\emph{Proceedings of the 33rd Annual ACM Symposium on
  User Interface Software and Technology}}. \bibinfo{pages}{126--139}.
\newblock


\end{thebibliography}
\end{document}